\documentclass[useAMS,usenatbib]{mn2e}
\usepackage{epsfig}
\usepackage{amsmath}

\newcommand{\be}{\begin{equation}}
\newcommand{\beq}{\begin{equation}}
\newcommand{\ba}{\begin{eqnarray}}
\newcommand{\ee}{\end{equation}}
\newcommand{\eeq}{\end{equation}}
\newcommand{\ea}{\end{eqnarray}}

\newcommand{\apj}{ApJ}
\newcommand{\apjl}{ApJL}
\newcommand{\mnras}{MNRAS}
\newcommand{\aj}{AJ}
\newcommand{\apjs}{ApJS}
\newcommand{\nat}{{\it Nature}}

\def\lsim{~\rlap{$<$}{\lower 1.0ex\hbox{$\sim$}}}

\def\gsim{~\rlap{$>$}{\lower 1.0ex\hbox{$\sim$}}}

\title[Correlation Between High Redshift Star Formation and 21cm
Emission]{The Correlation Between Star Formation and 21cm
Emission During the Reionization Epoch}

\author[Wyithe, Loeb \& Schmidt]{J. Stuart B. Wyithe$^1$, Abraham Loeb$^2$ \& Brian P. Schmidt$^3$\\$^1$
School of Physics, University of Melbourne, Parkville, Victoria,
Australia\\$^2$ Harvard-Smithsonian Center for Astrophysics, 60 Garden St.,
Cambridge, MA 02138\\$^3$ The Research School of Astronomy and Astrophysics, Weston Creek, ACT, Australia\\Email: swyithe@physics.unimelb.edu.au,
loeb@cfa.harvard.edu, brian@mso.anu.edu.au}

\begin{document}


\maketitle

\label{firstpage}
\begin{abstract}
 
Reionization is thought to be dominated by low mass galaxies, while direct
observations of resolved galaxies probe only the most massive, rarest
objects. The cross-correlation between fluctuations in the surface
brightness of the cumulative Ly$\alpha$ emission (which serves as a proxy
for the star formation rate) and the redshifted 21cm signal from neutral
hydrogen in the intergalactic medium (IGM), will directly probe the causal
link between the production of ionizing photons in galaxies and the
reionization of the IGM. We discuss the prospects for detecting this
cross-correlation for unresolved galaxies. We find that on angular scales $\la10^\prime$ detection will be
practical using widefield near-IR imaging from space in combination with the forthcoming Mileura Widefield Array - Low Frequency Demonstrator. When redshifted 21cm
observations of the neutral IGM are combined with space-based near-IR
imaging of Ly$\alpha$ emission, the detection on angular scales $\la3^\prime$
 will be limited by the sensitivity of the 21cm signal,
even when a small aperture optical telescope ($\sim2$m) and a moderate field of view
($\sim10$ square degrees) are used. On scales $\ga3^\prime$, the measurement of
cross-correlation will be limited by the accuracy of the foreground sky subtraction.

\end{abstract}

\begin{keywords}
cosmology: diffuse radiation, large scale structure, theory -- galaxies:
high redshift, intergalactic medium
\end{keywords}
 
\section{Introduction}

The primary goals for studies of the reionization epoch are to determine
the nature of the first generation of galaxies, and to observe the causal
link between these galaxies and the ionization state of the intergalactic
medium (IGM). At the current time, direct observations of resolved galaxies
probe only the most massive, rarest objects (Stark, Loeb \& Ellis 2007, and
references therein). It has been shown that these massive galaxies should
correlate with the redshifted 21cm signal from diffuse neutral hydrogen in
the IGM prior to the completion of reionization owing to the biased galaxy
formation in over-dense regions (Wyithe \& Loeb 2007; Furlanetto \& Lidz
2007). However, these massive galaxies are not responsible for the bulk of
the ionizing photons that reionized the IGM. Rather, reionization was
dominated by low mass galaxies, with luminosities below current detection
thresholds (Ellis 2007, and references therein). The emission of these
unresolved galaxies should therefore also be correlated with the ionization
of the IGM, and by extension, with the redshifted 21cm signal. In this paper we
suggest that the cross-correlation between the luminosity density of
unresolved Ly$\alpha$ emission and the redshifted 21cm intensity will
directly probe the connection between the reionization of the IGM and the
star formation rate (and hence the production of ionizing photons). We
compute the expected amplitude of this cross-correlation, and discuss the
prospects for its detection.

Star formation at high redshift has been studied using fluctuations in
unresolved near-IR broad-band emission (e.g. Kashlinsky et
al.~2005). Since the fluctuations from star formation at high redshift
are superimposed on fluctuations from foreground galaxies at low
redshift, these measurements have required subtraction of a model for
the fluctuating foreground component. In this paper we discuss
removal of the foreground fluctuations statistically using the fact that
these are uncorrelated with the redshifted 21cm emission. The
measurement of 21cm emission is also subject to a fluctuating foreground,
which will be correlated with the foreground in the Ly$\alpha$
observations. However it is proposed as part of upcoming 21cm
experiments, that the redshifted 21cm foreground be removed using the
smoothness of the spectrum of foreground sources, which will be
compared with the rapid frequency fluctuations of the 21cm signal
(Morales et al.~2006). This subtraction method will reveal the
narrow-band 21cm fluctuations, but will not allow detection of
broad-band 21cm fluctuations. Therefore, rather than consider
fluctuations in broad-band flux from high redshift star formation, in
this paper we instead discuss narrow-band near-IR observations. The
fluctuations in flux within narrow band observations would be
dominated by the Ly$\alpha$ line of galaxies in a narrow redshift
interval, and would therefore be the appropriate choice for detecting
the cross-correlation between the signals.

Any model for the reionization of the IGM must describe the relation
between the emission of ionizing photons by stars in galaxies and the
ionization state of the intergalactic gas. This relation is non-trivial as
it depends on various internal parameters (which may vary with galaxy
mass), such as the fraction of the gas within galaxies that is converted
into stars and accreting black holes, the spectrum of the ionizing
radiation, and the escape fraction of ionizing photons from the surrounding
interstellar medium as well as the galactic halo and its immediate infall
region (see Loeb 2006 for a review). The relation also depends on
intergalactic physics.  In regions of the IGM that are overdense, galaxies
will be over-abundant because small-scale fluctuations need to be of lower
amplitude to form a galaxy when embedded in a larger-scale overdensity (Mo
\& White~1996). On the other hand, the increase in the recombination rate
in over-dense regions counteracts this {\it galaxy bias}. The process of
reionization also contains several layers of feedback.  Radiative feedback
heats the IGM and results in the suppression of low-mass galaxy formation
(Efstathiou, 1992; Thoul \& Weinberg~1996; Quinn et al.~1996; Dijkstra et
al.~2004). This delays the completion of reionization by lowering the local
star formation rate, but the effect is counteracted in over-dense regions
by the biased formation of massive galaxies. Most models predict that the
sum of these effects is dominated by galaxy bias, and that as a result
over-dense regions are reionized first.  It follows that the
cross-correlation between the star formation rate density and redshifted
21cm emission should be negative, as has been suggested for the
cross-correlation between massive galaxies and redshifted 21cm emission
(Wyithe \& Loeb~2007; Furlanetto \& Lidz~2007).

A measurement of the expected anti-correlation between the local star
formation rate and the ionization state of the IGM, would provide crucial
evidence in favor of the stellar UV reionization model over alternative
models in which reionization resulted from decaying particles (Hansen \&
Haiman 2004; Bierman \& Kusenko 2006; Kasuya \& Kawasaki 2007; Ripamonti et
al. 2007) or from a more diffuse X-ray background (Madau et al. 2004;
Ricotti et al. 2005).  In this paper we examine the feasibility of making
this important measurement based on a simple illustrative model for stellar
reionization, described in \S 2. We then derive the cross-correlation
between the star formation rate and 21cm emission in \S~3, before
discussing the prospects for its detection in \S~4. Throughout the paper we
adopt the set of cosmological parameters determined by {\it WMAP} (Spergel
et al. 2006) for a flat $\Lambda$CDM universe.

\section{Density Dependent model of reionization}
\label{model}

In this paper we compute the relation between the local dark matter
overdensity and the brightness temperature of redshifted 21cm emission
based on the model described in Wyithe \& Loeb~(2006). Here we summarize
the main features of the model and refer the reader to that paper for more
details.

The evolution of the ionization fraction by mass $Q_{\delta,R}$ of a
particular region of scale $R$ with overdensity $\delta$ (at observed
redshift $z_{\rm obs}$) may be written as
\begin{eqnarray}
\label{history}
\nonumber
\frac{dQ_{\delta,R}}{dt} &=& \frac{N_{\rm ion}}{0.76}\left[Q_{\delta,R} \frac{dF_{\rm col}(\delta,R,z,M_{\rm ion})}{dt} \right.\\
\nonumber
&&\hspace{5mm}+ \left.\left(1-Q_{\delta,R}\right)\frac{dF_{\rm col}(\delta,R,z,M_{\rm min})}{dt}\right]\\
&-&\alpha_{\rm B}Cn_{\rm H}^0\left(1+\delta\frac{D(z)}{D(z_{\rm obs})}\right) \left(1+z\right)^3Q_{\delta,R},
\end{eqnarray}
where $N_{\rm ion}$ is the number of photons entering the IGM per baryon in
galaxies, $\alpha_{\rm B}$ is the case-B recombination co-efficient, $C$ is
the clumping factor (which we assume, for simplicity, to be constant), and
$D(z)$ is the growth factor between redshift $z$ and the present time. The
production rate of ionizing photons in neutral regions is assumed to be
proportional to the collapsed fraction $F_{\rm col}$ of mass in halos above
the minimum threshold mass for star formation ($M_{\rm min}$), while in
ionized regions the minimum halo mass is limited by the Jeans mass in an
ionized IGM ($M_{\rm ion}$). We assume $M_{\rm min}$ to correspond to a
virial temperature of $10^4$K, representing the hydrogen cooling threshold,
and $M_{\rm ion}$ to correspond to a virial temperature of $10^5$K,
representing the mass below which infall is suppressed from an ionized IGM
(Dijkstra et al.~2004). In a region of comoving radius $R$ and mean
overdensity $\delta(z)=\delta D(z)/D(z_{\rm obs})$ [specified at redshift $z$ instead of the usual
$z=0$], the relevant collapsed fraction is obtained from the extended
Press-Schechter~(1974) model (Bond et al.~1991) as
\begin{equation}
F_{\rm col}(\delta,R,z) = \mbox{erfc}{\left(\frac{\delta_{\rm
c}-\delta(z)}{\sqrt{2\left(\left[\sigma_{\rm
gal}\right]^2-\left[\sigma(R)\right]^2\right)}}\right)},
\end{equation}
where $\mbox{erfc}(x)$ is the error function, $\sigma(R)$ is the variance
of the density field smoothed on a scale
$R$, and $\sigma_{\rm gal}$ is the variance of the density field smoothed
on a scale $R_{\rm gal}$, corresponding to a mass scale of $M_{\rm min}$ or $M_{\rm ion}$ (both evaluated at redshift $z$ rather than at $z=0$).  In
this expression, the critical linear overdensity for the collapse of a
spherical top-hat density perturbation is $\delta_c\approx
1.69$.

Equation~(\ref{history}) may be integrated as a function of
$\delta$.  At a specified redshift, this yields the filling fraction
of ionized regions within the IGM on various scales $R$ as a function
of overdensity. We may then also calculate the corresponding 21cm
brightness temperature contrast 
\begin{equation} T(\delta,R) =
22\mbox{mK}(1-Q_{\delta,R})\left(\frac{1+z}{7.5}\right)^{-0.5}\left(1+\frac{4}{3}\delta\right),
\end{equation} 
where the pre-factor of 4/3 on the overdensity refers
to the spherically averaged enhancement of the brightness temperature
due to peculiar velocities in over-dense regions (Bharadwaj \&
Ali~2005; Barkana \& Loeb~2005).

\section{The Ly$\alpha$ luminosity density}

The density dependent model described in the previous section may be used to estimate the cross-correlation between star formation rate and the ionization state of the IGM. In this section, we begin by computing the star formation rate. Then, in subsequent sections we estimate the auto-correlation functions for both star formation rate and 21cm brightness temperature, as well as the cross-correlation between star formation rate and 21cm brightness temperature.

The UV-luminosity of galaxies is largest during periods of active star formation. In the dense environments within the high redshift inter-stellar medium the density of neutral hydrogen can be substantial, resulting in absorption of the majority of the UV photons produced. Recombinations in the ionized hydrogen then in turn produce Ly$\alpha$ photons. The Ly$\alpha$ emission from high redshift galaxies is therefore powered by concurrent star formation. In this paper we assume Ly$\alpha$ emissivity to be a proxy for the
star formation rate, and so begin by computing the luminosity
density of Ly$\alpha$ photons. Given an ionizing photon production
rate
\begin{eqnarray}
\nonumber
\label{Idot}
\log_{10}\left(\frac{\dot{I}}{\mbox{sec}^{-1}}\right) &=& \\
&&\hspace{-25mm}53.8 + \log_{10}\left(\frac{\dot{M}}{M_\odot\mbox{yr}^{-1}}\right)-0.0029\left(9+\log_{10}(Z)\right)^{2.5},
\end{eqnarray}
where $\dot{M}$ is the star formation rate per comoving Mpc$^3$, and $Z$ the metalicity of a stellar population with a Salpeter initial mass function,
the luminosity of Ly$\alpha$ entering the IGM is 
\begin{equation}
\label{SFR}
\Gamma = 2h_{\rm p}\frac{\nu_{\alpha}}{3}(1-f_{\rm esc})\mathcal{T}\dot{I},
\end{equation}
where $f_{\rm esc}$ is the escape fraction of ionizing photons,
$h_{\rm p}$ is Planck's constant and $\nu_{\alpha}$ is the frequency of
the Ly$\alpha$ transition. The transmission of Ly$\alpha$ photons through the IGM ($\mathcal{T}$) is less than unity and is discussed below. In the above expressions we evaluate the
star formation rate within a region of comoving radius $R$ as
\begin{eqnarray}
\nonumber
\dot{M}&=& f_{\rm star}\frac{\Omega_{\rm b}}{\Omega_{m}}\rho_{\rm m}(1+\delta)\left((1-Q_{\delta,R})\frac{dF_{\rm col}(\delta,R,M_{\rm min})}{dt}\right.\\
&&\hspace{28mm}+\left.Q_{\delta,R}\frac{dF_{\rm col}(\delta,R,M_{\rm ion})}{dt}\right).
\end{eqnarray}
Here $f_{\rm star}$ is the star formation efficiency, $\Omega_{\rm b}$ and
$\Omega_{\rm m}$ are the density parameters in matter and baryons, and
$\rho_{\rm m}$ is the average comoving mass-density in the Universe.

It is possible that the mass-function of stars in Ly$\alpha$ emitting
galaxies is top-heavy, in which case the Ly$\alpha$ luminosity could
be an order of magnitude greater than suggested by
equations~(\ref{Idot}-\ref{SFR}). Indeed Dijkstra \& Wyithe~(2007)
have noted that this must be the case due to the the large observed equivalent widths in known
Ly$\alpha$ emitters, and the small
value of Ly$\alpha$ transmission through the IGM (Dijkstra, Lidz \&
Wyithe~2007). However Dijkstra \& Wyithe~(2007) also argue that while
top-heavy star formation must be present in many high redshift Ly$\alpha$ emitters, in order to be consistent
with additional observations the top-heavy formation phase must last for less than 10\% of the star formation time-scale in individual galaxies. As a result, Dijkstra \& Wyithe~(2007) find that
the total Ly$\alpha$ emission is dominated by a normal stellar
population when averaged over the full star formation history of galaxies at $z\sim6$.

\subsection{The transmission of Ly$\alpha$ photons through the IGM}

Due to the strength of the Ly$\alpha$ resonance, a significant
fraction of Ly$\alpha$ flux is absorbed in the infalling IGM
surrounding a galaxy (Dijkstra, Lidz \& Wyithe~2007). The quantity
$\mathcal{T}$ in equation~(\ref{SFR}) is the transmission of
Ly$\alpha$ photons through the IGM, and corresponds to the fraction of
Ly$\alpha$ photons leaving the galaxy that propagate to an
observer. We assume the absorption of Ly$\alpha$ photons in the IGM to
be dominated by neutral hydrogen, with a negligible contribution from
dust (due to the low metalicity of the high redshift IGM). We also
ignore absorption of Ly$\alpha$ photons by dust within the galaxy due
to the low metalicity of high redshift stellar populations. In biased
models of reionization, over dense regions are reionized first due to
their being regions of greater than average star formation. Thus
over-dense regions produce positive fluctuations in the luminosity
density of galactic Ly$\alpha$ emission. Conversely, neutral hydrogen
is located preferentially in under dense regions, which will therefore
be sites of lower Ly$\alpha$ transmission. As a result, the variable
transmission of Ly$\alpha$ photons could serve to increase the
clustering of Ly$\alpha$ galaxies (McQuinn et al.~2007), and hence to
also increase the amplitude of fluctuations in the density of
Ly$\alpha$ emission. Recently, Dijkstra, Lidz \& Wyithe~(2007) have
conducted a detailed investigation of the Ly$\alpha$ absorption
properties of the IGM surrounding a Ly$\alpha$ emitting galaxy. This
work concluded that the ionized IGM introduces significant absorption,
and that as a result the Ly$\alpha$ transmission is only weakly
dependent on the ionization state of the IGM. In particular,
Ly$\alpha$ flux from a galaxy embedded in an HII region rather than in
a reionized IGM will be subject to only a small amount of additional
absorption due to the damping wing of the Ly$\alpha$ resonance. Rather
than introduce a complex model for transmission, in this paper we
instead assume the transmission to have the same value for all
galaxies, and to be independent of overdensity. As a result we may
underestimate the amplitude of Ly$\alpha$ fluctuations. The increased
fluctuations introduced by variable transmission would increase the
amplitude of the cross-correlation signal between Ly$\alpha$ and
redshifted 21cm emission. By assuming constant transmission we
therefore arrive at conservative estimates for the detectability of
the cross-correlation signal.

\begin{figure*}
\includegraphics[width=15cm]{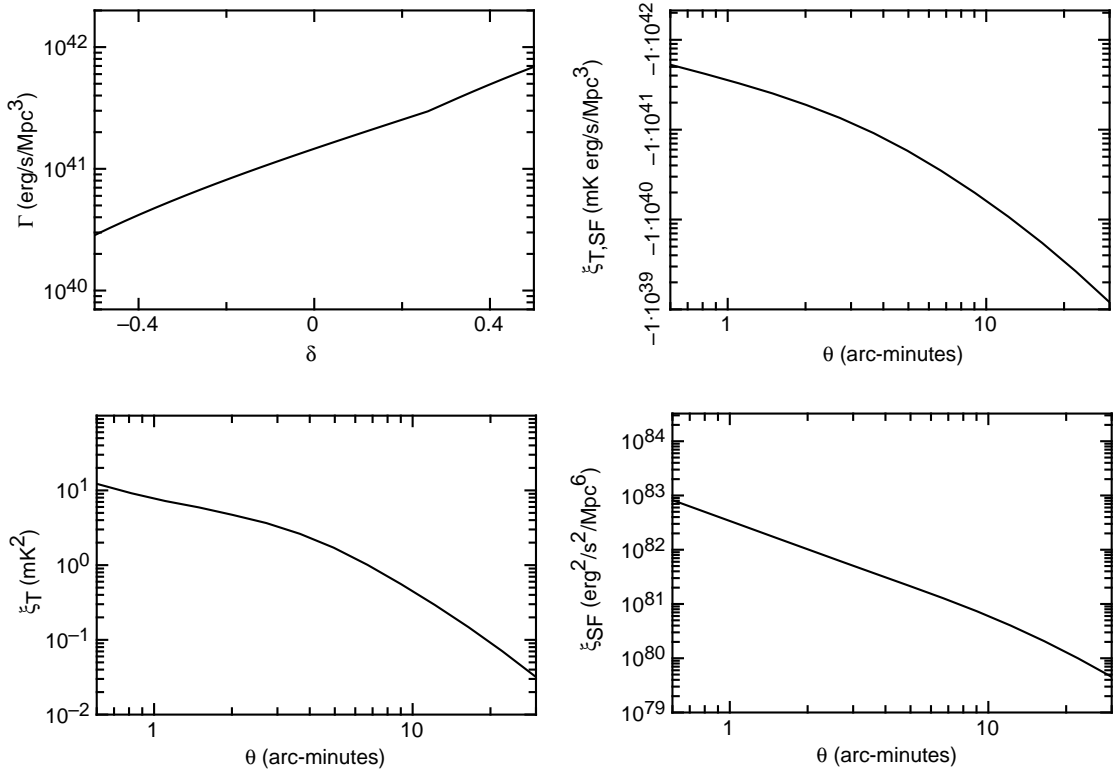} 
\caption{ \textit{Upper Left:} The luminosity density (erg per second per
comoving Mpc) in the Ly$\alpha$ line as a function of the large scale
overdensity ($\delta$).  \textit{Upper Right:} The cross-correlation
function of the luminosity density in the Ly$\alpha$ line, with the 21cm
brightness temperature contrast within spheres of observed radius
$\theta$. \textit{Lower Left:} The auto-correlation function of 21cm
brightness temperature within spheres of observed radius
$\theta$. \textit{Lower Right:} The auto-correlation function of
luminosity density in the Ly$\alpha$ line within spheres of observed radius
$\theta$. }
\label{fig1}
\end{figure*}

\subsection{Diffuse Ly$\alpha$ emission from the IGM}

In calculating the Ly$\alpha$ luminosity density we have neglected the
potential contribution from a re-combining IGM. We now show that this process provides a negligible contribution. To see this we note that at $z=7$, around 10\% of baryons are collapsing inside galaxies per
Hubble time, and that some fraction of these form stars ($\sim30\%$). For
every baryon taking part in star formation, around 4000 ionizing
photons are produced (e.g. Barkana \& Loeb~2001). Most ionizing photons do not escape the galaxy,
and each of these produces around 2/3 of a Ly$\alpha$ photon. Of the
Ly$\alpha$ photons produced, some ($\sim70\%$) will be absorbed in the
IGM surrounding the galaxy. Hence we find
$\sim4000\times0.1\times0.3\times(2/3)\times(1-0.7)$, or around 25 photons per baryon are produced
by galaxies during 1 Hubble time.  On the other-hand, at the redshift
of interest, the recombination rate per baryon is around
once per Hubble time, yielding of order 2/3 Ly$\alpha$ photons per
baryon per Hubble time from the diffuse IGM. This number is 1.5 orders
of magnitude smaller than the galactic Ly$\alpha$ emission. A more quantitative estimate of this ratio $R_{\rm Ly}$ is
\begin{equation}
R_{\rm Ly}\sim50\left(\frac{t_{\rm H}\frac{d\bar{F}_{\rm col}}{dt}}{0.1}\right)\left(\frac{(f_{\rm star}\mathcal{T})N_\gamma}{400}\right)\left(\frac{1+z}{10}\right)^{-\frac{3}{2}}\left(\frac{\bar{Q}}{0.5}\right)^{-1},
\end{equation}
where $\bar{Q}$ and $\bar{F}_{\rm col}$ are the average ionized fraction and collapsed fraction in the IGM respectively, and $N_\gamma$ is the number of ionizing photons produced per baryon incorporated into stars.

\section{Fiducial Model for Reionization}

In this paper we show results for the cross-correlation between
Ly$\alpha$ and 21cm emission for a model that reionizes the mean IGM
at $z=6$ (White et al.~2003). In this model we assume that
star formation proceeds in halos above the hydrogen cooling threshold
in neutral regions of IGM. In ionized regions of the IGM star formation
is assumed to be suppressed by radiative feedback (see
\S~\ref{model}). In what follows we present estimates of fluctuations
in flux due to sources at $z=7$, at which time the IGM is around 70\%
ionized in this model. We compute values for auto-correlation
functions, and the 21cm-Ly$\alpha$ cross-correlation function at
scales as small as 0.6$^\prime$. However our model begins to break
down on scales below $\sim1^\prime$, where at $z=7$, 10\% of regions
have already been reionized on this scale (Wyithe \& Morales~2007).

\section{Variation of Ly$\alpha$ emission with overdensity}

The top-left panel of Figure~\ref{fig1} shows the luminosity density (erg per second per comoving Mpc) in the Ly$\alpha$ line as a function of the large scale overdensity ($\delta$). To calculate the level of observed Ly$\alpha$ emission, we require an estimate of the product $f_{\rm star}\mathcal{T}$ (only the product of these parameters enters the observed luminosity). In a recent analysis Dijkstra, Wyithe \& Haiman~(2007) have used semi-analytic models to constrain this parameter using the observed luminosity function of Ly$\alpha$ emitting galaxies at $z=5.7$ and $z=6.5$. The constraint is sensitive to the life-time of the Ly$\alpha$ emission, but is expected to fall in the range $0.03\la f_{\rm star}\mathcal{T}\la0.1$. Here, and in the remainder of this paper we assume the product $\mathcal{T}f_{\rm star}=0.1$ when considering the properties of high redshift Ly$\alpha$ emitters.

\section{Auto-correlation functions for star formation and 21cm emission}

Before discussing the cross-correlation of star formation rate (Ly$\alpha$
emission) with 21cm emission, we first compute each of the auto-correlation
functions individually.  On comoving scales $R$ larger than the characteristic bubble size ($\ga1^\prime$ at $z=7$ in our model), we are able to compute the auto-correlation
function [$\xi_T(\theta)$] of fluctuations in brightness temperature $T$
smoothed with top-hat windows of angular radius $\theta=R/D_{\rm A}(z)$,
\begin{eqnarray}
\nonumber
\label{autoT}
\xi_T(\theta) &=& \langle \left(T-\langle T\rangle\right)^{2}\rangle^{1/2}\\
 &=& \left[\frac{1}{\sqrt{2\pi}\sigma(R)}\int d\delta \left(T(\delta)-\langle T\rangle\right)^2e^{-\frac{\delta^2}{2\sigma(R)^2}} \right]^{\frac{1}{2}}.
\end{eqnarray}
Here
\begin{equation}
\langle T\rangle = \frac{1}{\sqrt{2\pi}\sigma(R)}\int d\delta~ T(\delta)e^{-\frac{\delta^2}{2\sigma(R)^2}},
\end{equation}
and $\theta=R/D_{\rm A}$ where $D_{\rm A}$ is the angular diameter distance.
The auto-correlation function of 21cm brightness temperature within spheres of observed radius $\theta$ is plotted in the lower-left panel of Figure~\ref{fig1}.

We also compute the auto-correlation function
[$\xi_{\rm SF}(\theta)$] of fluctuations in Ly$\alpha$ emission $\Gamma$
smoothed with top-hat windows of angular radius $\theta=R/D_{\rm A}(z)$,
\begin{eqnarray}
\nonumber
\xi_{\rm SF}(\theta) &=& \langle \left(\Gamma-\langle\Gamma\rangle)^{2}\right\rangle^{1/2}\\
 &=& \left[\frac{1}{\sqrt{2\pi}\sigma(R)}\int d\delta \left(\Gamma-\langle\Gamma\rangle\right)^2e^{-\frac{\delta^2}{2\sigma(R)^2}} \right]^{\frac{1}{2}},
\end{eqnarray}
where 
\begin{equation}
\langle \Gamma \rangle = \frac{1}{\sqrt{2\pi}\sigma(R)}\int d\delta~ \Gamma(\delta)e^{-\frac{\delta^2}{2\sigma(R)^2}}.
\end{equation}
The auto-correlation function of luminosity density in the Ly$\alpha$ line within spheres of observed radius $\theta$ is shown in the lower-right panel of Figure~\ref{fig1}.

\section{The cross-correlation function between Ly$\alpha$ and 21cm emission}

The properties of the galaxy population are expected to correlate with
the level of redshifted 21cm emission. These properties depend on the
overdensity of the IGM whose typical fluctuation level is a function
of scale. As a result, the amplitude of the correlation between
fluctuations in Ly$\alpha$ emission ($\Gamma-\langle\Gamma\rangle$)
and fluctuations in 21cm brightness temperature contrast ($T-\langle
T\rangle$) will therefore also be dependent on angular scale. On a
comoving scale $R$ larger than the characteristic bubble radius, we
are able to compute the cross-correlation function between Ly$\alpha$
and 21cm emission \begin{eqnarray} \nonumber \xi_{\rm T,{\rm
SF}}(\theta)&=&\langle(\Gamma-\langle\Gamma\rangle)\times(T-\langle
T\rangle)\rangle\\ &&\hspace{-15mm}=\frac{1}{\sqrt{2\pi}\sigma(R)}\int
d\delta\left((\Gamma-\langle\Gamma\rangle)\times(T-\langle
T\rangle)\right)e^\frac{-\delta^2}{2\sigma(R)^2} \end{eqnarray} for
the IGM smoothed on various angular scales.  The resulting
cross-correlation function of the luminosity density in the Ly$\alpha$
line, with the 21cm brightness temperature contrast within spheres of
observed radius $\theta$ is shown in the top-right panel of
Figure~\ref{fig1}. The sign of this cross-correlation is negative,
indicating an anti-correlation between star formation and 21cm
emission. This anti-correlation arises as a result of the higher star formation rates generated due to galaxy bias in overdense regions, which are therefore reionized first. The amplitude of the cross-correlation
decreases towards large scales.

\section{Detectability of the cross-correlation signal}

\begin{figure*}
\includegraphics[width=11cm]{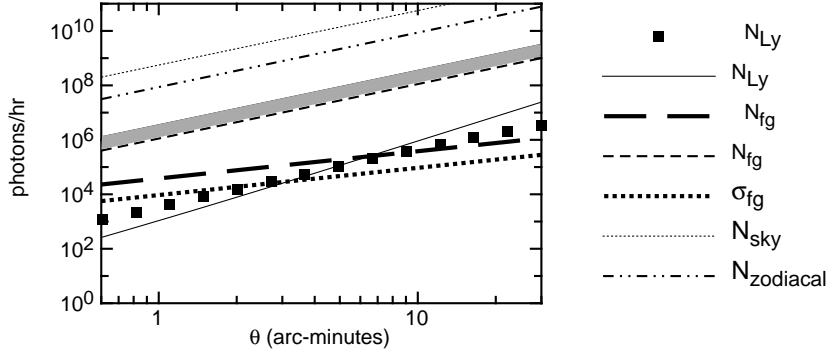} 
\caption{ The observed fluxes and fluctuations in observed fluxes in units
of photons per hour. Calculation of the flux was performed assuming a 2m
telescope and a 1 hour integration.  The large black dots and thin solid
line refer respectively to the fluctuation level ($\Delta \dot{N}_{\rm
Ly}$) and the total level ($\dot{N}_{\rm Ly}$) of Ly$\alpha$ emission
within a spherical region of observed radius $\theta$.  The thin dashed
curve corresponds to the flux ($\dot{N}_{\rm fg}$) in a $100$\AA~ band due to foregrounds at the
wavelength of the observed Ly$\alpha$ emission. For comparison we also plot
the measured extra-galactic foreground at $8000$\AA~  as the grey band. 
The level of fluctuations ($\Delta\dot{N}_{\rm fg}$) in $\dot{N}_{\rm fg}$ among different lines of sight
due to Poisson noise in the number of galaxies contributing to the
background is shown as the thick dashed line. The levels of sky-glow ($\dot{N}_{\rm sky}$) and zodiacal light
($\dot{N}_{\rm zodiacal}$) are shown by the thin dotted and dot-dot-dashed lines respectively. Finally the thick
dotted line shows the Poisson noise ($\sigma_{\rm fg}=\sqrt{\dot{N}_{\rm fg}+\dot{N}_{\rm Ly} + \dot{N}_{\rm zodiacal}}$) in the number of
photons detected in a space-based observation (i.e. no sky-glow) per region of radius $\theta$.}
\label{fig2}
\end{figure*} 

In the remainder of this paper we discuss detection of the predicted
cross-correlation between Ly$\alpha$ and the 21cm emission. We
begin with the Ly$\alpha$ signal and extra-galactic foreground, which
we assume are measured in a wide-field near-IR survey through a narrow-band filter centered on the redshifted Ly$\alpha$ wavelength. We then
discuss the sensitivity of planned low-frequency arrays to the
redshifted 21cm signal, before describing prospects for detection of
the predicted cross-correlation between Ly$\alpha$ and redshifted 21cm
emission using a range of current and future observational facilities.

\subsection{The Ly$\alpha$ flux}

Equation~(\ref{autoT}) can be used to compute the fluctuations in Ly$\alpha$ luminosity from spherical regions subtending an angle $\theta$,
\begin{equation}
\Delta L_{\rm Ly}=\left(\xi_{\rm SF}\right)^{1/2}\frac{4\pi(\theta D_{\rm A})^3}{3},
\end{equation}
while the corresponding total luminosity follows from equation~(\ref{SFR})
\begin{equation}
L_{\rm Ly}=\langle\Gamma\rangle\frac{4\pi(\theta D_{\rm A})^3}{3}.
\end{equation}

For a telescope of diameter $d$, the fluctuations in the observed photon count are
\begin{equation}
\Delta \dot{N}_{\rm Ly} = \pi\left(\frac{d}{2}\right)^2\frac{\Delta L_{\rm Ly}}{4\pi D_{\rm L}^2}\frac{1}{h_{\rm p}\nu_{\rm obs}},
\end{equation}
where $D_{\rm L}$ is the luminosity distance and $\nu_{\rm obs}$ is the observed frequency of the Ly$\alpha$ photons. Similarly, the total flux in Ly$\alpha$ photons is 
\begin{equation}
\dot{N}_{\rm Ly} = \pi\left(\frac{d}{2}\right)^2\frac{L_{\rm Ly}}{4\pi D_{\rm L}^2}\frac{1}{h_{\rm p}\nu_{\rm obs}}.
\end{equation}
Figure~\ref{fig2} shows the observed fluxes and fluctuations in observed fluxes in units of photons per hour. In Figure~\ref{fig2} we assumed a 2m telescope with a 1 hour integration time. The large black dots and thin solid line refer  respectively to the fluctuation level ($\Delta \dot{N}_{\rm Ly}$) and the total level ($\dot{N}_{\rm Ly}$) of Ly$\alpha$ emission within a spherical region of observed radius $\theta$.

\subsection{Foreground emission}

Observations through a narrow filter will detect fluctuations
in Ly$\alpha$ emission superimposed on fluctuations in the extra-galactic foreground. To estimate the importance of the foreground with respect to measurement of the cross-correlation we therefore need to estimate both the total foreground flux ($F$), and the fluctuations in total flux that enters the detector from a
cone of angular radius $\theta$ at a frequency $\nu_{\rm obs}$. For
our purposes it is sufficient to estimate the foreground flux using the
following simple model
\begin{eqnarray}
\nonumber
F(\theta,\nu_{\rm obs}) &=&\\
&&\hspace{-15mm} \int_0^{z_{\rm Ly}} dz \pi[D_{\rm A}(z)\theta]^2 \frac{cdt}{dz}\left.\frac{d^2E}{dVd\nu}\right|_{\nu=\nu_{\rm obs}(1+z)}\frac{(1+z)}{4\pi D_{\rm L}^2},
\end{eqnarray}
where 
\begin{equation}
\frac{d^2E}{dVd\nu}=f_\star\frac{\Omega_{\rm b}}{\Omega_{\rm m}}\rho_{\rm m}(1+z)^3\frac{dF_{\rm col}}{dt}\frac{d^2E}{d\dot{M}d\nu}
\end{equation}
is the luminosity density. In the latter expression,
$\frac{d^2E}{d\dot{M}d\nu}$ is the luminosity produced at frequency
$\nu$ per unit star formation rate. We assume a 1/20th solar metalicity population with a Scalo~(1998) mass-function, and use the stellar population model of Leitherer et al.~(1999) to compute the spectrum of a continuously star forming galaxy\footnote{Model spectra of star forming galaxies obtained from http://www.stsci.edu/science/starburst99/.}. For a narrow filter of width $\Delta\nu_{\rm Ly}$, the flux can then be converted
into a photon detection rate
\begin{equation}
\dot{N}_{\rm fg} = \pi\left(\frac{d}{2}\right)^2\frac{F(\theta,\nu_{\rm obs})}{h_{\rm p}\nu_{\rm obs}}\Delta\nu_{\rm Ly},
\end{equation}
where $h_{\rm p}$ is Planck's constant. The thin dashed curve in
Figure~\ref{fig2} corresponds to the flux in a 100\AA~ band due to
foregrounds at the wavelength of the observed Ly$\alpha$ emission from
galaxies at $z=7$. This model can be compared to the measured
extra-galactic foreground at 8000\AA. Bernstein, Freedman \& Madore~(2002)
find flux at a level of (1.2-3)$\times10^{-9}$erg/s/cm$^2$/\AA/Sr. 
In the units of Figure~\ref{fig2}, this observation is shown as the grey
band. Our model estimate lies on the lower boundary of the measured range for the observed
foreground.

There will be fluctuations ($\Delta \dot{N}_{\rm fg}$) in $\dot{N}_{\rm fg}$ among different lines
of sight due to Poisson noise in the number of galaxies contributing
to the foreground. The level of fluctuations is given by
\begin{eqnarray} 
\label{fgfluct}
\nonumber
\frac{\Delta\dot{N}_{\rm fg}}{\dot{N}_{\rm fg}} &=&\\
&&\hspace{-17mm}\frac{\sqrt{\int_0^{z_{\rm Ly}} dz \int_{M_{\rm ion}}^{M_{\rm lim}(z)}
dM \epsilon_{\rm lt} \frac{dn}{dM} \pi \left[D_{\rm D}(z)\theta\right]^2 \frac{cdt}{dz}
\left[\dot{\mathcal{N}}(M,z)\right]^2 } } {\int_0^{z_{\rm Ly}} dz
\int_{M_{\rm ion}}^{M_{\rm lim}(z)} dM \epsilon_{\rm lt} \frac{dn}{dM} \pi \left[D_{\rm
D}(z)\theta\right]^2 \frac{cdt}{dz} \dot{\mathcal{N}}(M,z)}, 
\end{eqnarray} 
where $\dot{\mathcal{N}}(M,z)$ is the observed flux from a galaxy of mass
$M$ at redshift $z$, and $\epsilon_{\rm lt}$ is the duty-cycle. The
presence of bright, resolved galaxies at low redshift increase the
fluctuations in the smoothed foreground. To reduce the amplitude of
fluctuations in the foreground, these resolved galaxies need to be
removed. To estimate the foreground fluctuations in the absence of resolved
galaxies, we therefore compute the flux corresponding to a photon limited
signal-to-noise ratio of 30, given an assumed telescope diameter,
integration time and filter width. As a function of redshift, we then
estimate the star formation rate ($\dot{M}_{\rm lim}$) corresponding to
this limiting flux. By assuming a star formation efficiency and lifetime
(we take $f_{\rm star}=0.1$ and $\epsilon_{\rm lt}=1$ respectively, which
is appropriate for the old stellar populations contributing to the
foreground), we estimate the limiting galaxy mass, $M_{\rm lim} =
\dot{M}_{\rm lim}t_{\rm lt}f_{\rm star}^{-1}\Omega_{\rm m}/\Omega_{\rm
b}$. Only galaxies whose masses are below this limit contribute to the
unresolved foreground.  The upper limits on the mass integration in
equation~(\ref{fgfluct}) therefore correspond to the minimum galaxy mass
that can be removed from the map as a resolved source prior to the
calculation of fluctuations.

\begin{figure*}
\includegraphics[width=10.5cm]{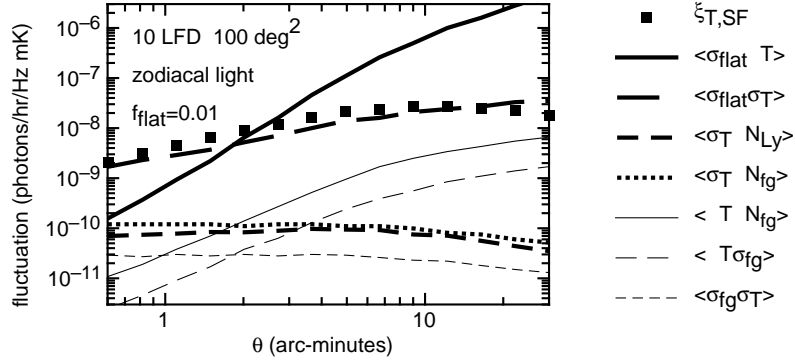} 
\caption{The signal and noise terms in equation~(\ref{CC}) as a function of $\theta$. This example assumed a Ly$\alpha$ survey with an area of $A=100$ square degrees using a 2 meter space-based telescope and 1hr integration per pointing, combined with a low-frequency array of collecting area 10 times the LFD with an integration time of 1000 hours. The near-IR observations were assumed to be flattened at the 1\% level.}
\label{fig3}
\end{figure*}

Figure~\ref{fig2} shows that the level of foreground flux (thin dashed
line) is larger than the flux due to galactic Ly$\alpha$ emission at $z\sim7$ (thin
solid line) by several orders of magnitude. However the absolute levels of
fluctuation are similar (thick dashed and large dotted lines), particularly
at larger angular scales\footnote{Note that the flux levels of the
foreground and the Ly$\alpha$ signal shown in Figure~\ref{fig2} have
different power-law dependences on $\theta$. This is because the
foreground has been calculated in a cone, while the Ly$\alpha$ emission has
been calculated in spheres.}. The fluctuations in the foreground and
Ly$\alpha$ emission have different angular dependencies because the
foreground fluctuations are dominated by Poisson noise, while the
Ly$\alpha$ fluctuations are a result of the biased star formation rates
toward over-dense regions.

In addition to the extra-galactic foreground generated by stellar continuum,
we expect a fluctuating foreground due to emission lines at frequencies
blueward of Ly$\alpha$, from sources located at redshifts below the
Ly$\alpha$ emitting galaxies. Like the Ly$\alpha$ fluctuations, these will
have large relative amplitudes due to the narrow redshift bin in which the
sources contribute to the foreground. However unlike the Ly$\alpha$
fluctuations this foreground of line emission will not correlate with the
redshifted 21cm emission. Since, as we show below, the fluctuating
foreground does not provide the limiting factor in detection of the
correlation between Ly$\alpha$ emission and 21cm intensity, we therefore
neglect the contribution of low redshift emission line sources to the
extra-galactic foreground in the remainder of this paper.

However images will be contaminated with additional foregrounds due to
zodiacal light, and, for ground based observation, due to atmospheric
sky glow in addition to the extragalactic foreground. In
Figure~\ref{fig2} we show the level of sky-glow based on the Keck
skyglow spectrum ($\dot{N}_{\rm sky}$) as the thin dotted line, and
the level of zodiacal light measured at $9000$\AA~ ($\dot{N}_{\rm
zodiacal}$) towards the ecliptic pole (Leinert et al.~1997) as the
dot-dot-dashed line.  Zodiacal light is around 2 orders of magnitude
larger, and atmospheric skyglow around 3 orders of magnitude larger
than the extragalactic foreground.

In order to detect the fluctuations in the correlation between
Ly$\alpha$ and 21cm emission, the observational noise in the
foreground must be smaller than the fluctuations being measured.  The
thick dotted line in Figure~\ref{fig2} shows the Poisson noise
($\sigma_{\rm fg}=\sqrt{\dot{N}_{\rm fg}+\dot{N}_{\rm Ly} + \dot{N}_{\rm
zodiacal}}$) in the number of photons detected per region of radius
$\theta$ (including zodiacal light but not atmospheric sky-glow). For the example shown the Poisson noise
is comparable to the size of fluctuations in Ly$\alpha$ emission.

\subsection{Near-IR field flatness}

Finally, we mention one further source of fluctuations that could mask
the fluctuations in Ly$\alpha$ emission. Variability of the
foregrounds in time and space can be removed through dithering
techniques to produce a foreground free, and flattened field
containing only the differential fluctuations in the signal (plus
extragalactic foregrounds). However the field can only be flattened to
a finite fractional level (e.g. 0.01), which we define to be $f_{\rm
flat}$.  Observations will therefore contain an additional fluctuating
term with an amplitude of $\sigma_{\rm flat}=f_{\rm
flat}(\dot{N}+\dot{N}_{\rm fg}+\dot{N}_{\rm sky}+\dot{N}_{\rm zodiacal})$. We will find that 
this term dominates the error budget on angular scales greater than a
few arc-minutes, and in the case of ground based observations, that it will prevent detection of the cross-correlation on those scales.

\subsection{Sensitivity to the 21cm signal}

In this section we
discuss the response of a phased array to the brightness temperature
contrast of the 21cm emission from the IGM. We define the error in brightness
temperature per synthesized beam
to be $\sigma_{\rm T}$. Assuming that calibration can be performed ideally,
and that redshifted 21cm foreground subtraction is perfect, the
root-mean-square fluctuations in brightness temperature are given by the
radiometer equation
\begin{equation}
\sigma_{\rm T} = \frac{\epsilon\lambda^2T_{\rm sys}}{A_{\rm
tot}\Omega_{\rm b}\sqrt{t_{\rm int}\Delta\nu_{21}}},
\end{equation}
where $\lambda$ is the wavelength, $T_{\rm sys}$ is the system
temperature, $A_{\rm tot}$ the collecting area, $\Omega_{\rm b}$ the
effective solid angle of the synthesized beam in radians, $t_{\rm
int}$ is the integration time, $\Delta\nu_{21}$ is the size of the
frequency bin, and $\epsilon$ is a constant that describes the overall
efficiency of the telescope. We optimistically adopt $\epsilon=1$ in this
paper. In units relevant for upcoming telescopes and at $\nu=200$MHz,
we find (Wyithe, Loeb \& Barnes~2005)
\begin{eqnarray}
\label{noiseeqn}
\nonumber
\sigma_{\rm T} &=& 7.5 \mbox{mK} \left(\frac{1.97}{C_{\rm beam}}\right) \left(\frac{A_{\rm tot}}{A_{\rm LFD}}\right)^{-1}\\
&\times&\left(\frac{\Delta\nu_{21}}{1\mbox{MHz}}\right)^{-1/2}\left(\frac{t_{\rm int}}{100\mbox{hr}}\right)^{-1/2}\left(\frac{\theta_{\rm beam}}{5^\prime}\right)^{-2}.
\end{eqnarray}
The label {\it LFD} corresponds to the Low-Frequency Demonstrator of the
Mileura Wide-Field Array (see
http://www.haystack.mit.edu/ast/arrays/mwa/site/index.html).  $A_{\rm LFD}$
is the collecting area of a phased array consisting of 500 tiles each with
16 cross-dipoles [the effective collecting area of an LFD tile with
$4\times4$ cross-dipole array with 1.07m spacing is $\sim17$--$19$m$^2$
between 100 and 200MHz (B. Correy, private communication)]. The system
temperature at 200MHz will be dominated by the sky and has a value $T_{\rm
sys}\sim250$K.  The size of the synthesized beam $\theta_{\rm beam}$ can be
regarded as the radius of a hypothetical top-hat beam, or as the variance
of a hypothetical Gaussian beam. The corresponding values of the constant
$C_{\rm beam}$ are 1 and 1.97 respectively.

\subsection{Estimate of signal-to noise ratio in detection of the cross-correlation}

\begin{figure*}
\includegraphics[width=17.5cm]{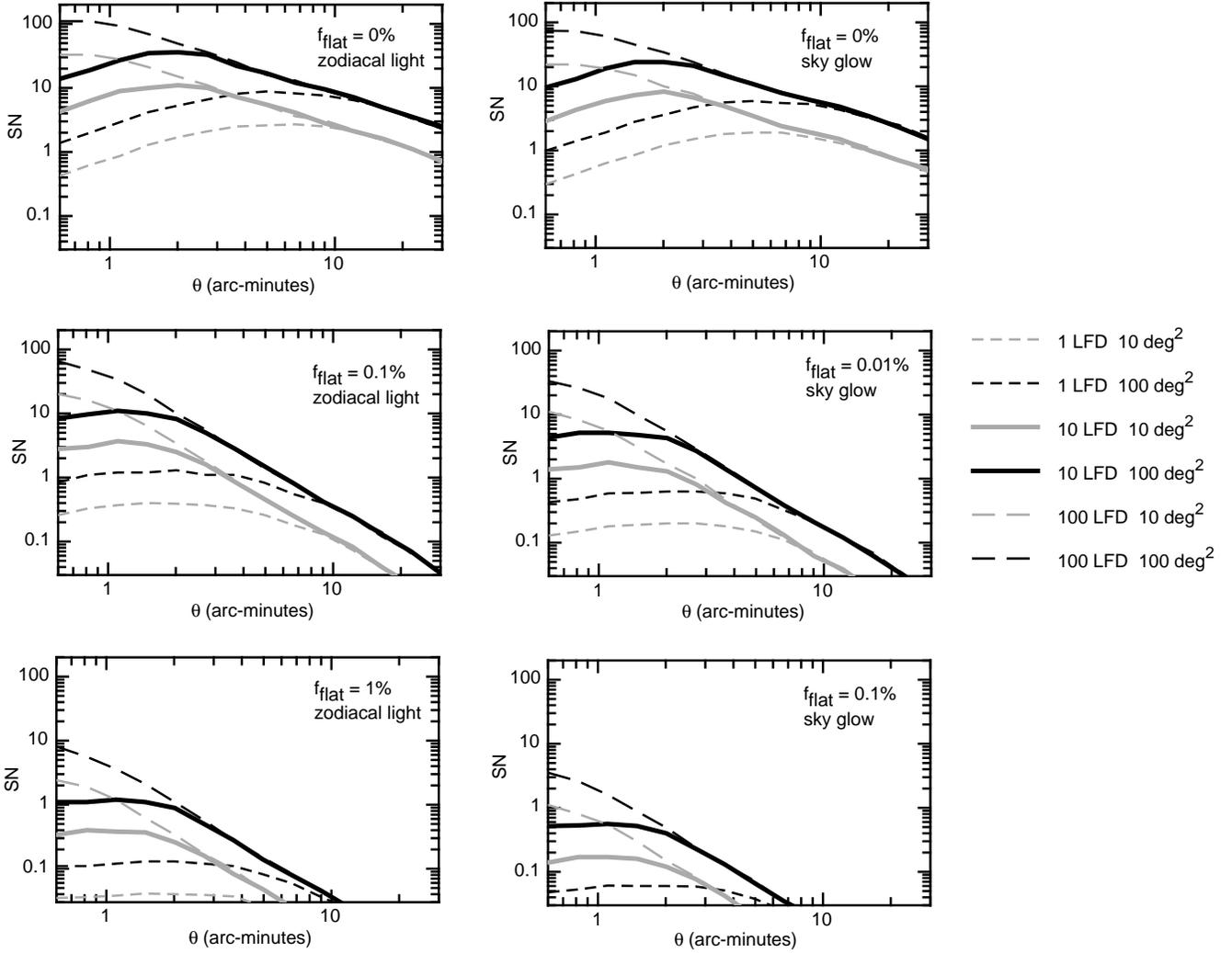}
\caption{ Signal to noise ratios as a function of angle. In each panel six
cases are shown, corresponding to Ly$\alpha$ surveys with areas of $A=10$
and 100 square degrees using a 2m telescope and a 1 hour integration;
combined with low-frequency arrays of collecting area corresponding to 1, 10
and 100 LFDs with an integration time 1000 hours. The left-hand panels correspond to
space based (i.e. no sky-glow, but including zodiacal light), and the right-hand panels to ground based near-IR observations (i.e. including
sky glow). The value of $f_{\rm flat}$ is listed in each
case. Note the assumed values for $f_{\rm flat}$ are an order of magnitude lower for ground based observations. }
\label{fig4}
\end{figure*}

The observed cross-correlation function ($\xi_{\rm Ly,T}$) is a combination of real fluctuations and noise, hence we can write 
\begin{eqnarray} 
\label{CC}
\nonumber
\xi_{\rm Ly,T}^{\rm obs} &=& \langle\left(\Delta \dot{N}_{\rm Ly}+\Delta\dot{N}+\sigma_{\rm fg}+\sigma_{\rm flat}\right)\left(\Delta T + \sigma_{\rm T} \right) \rangle \\
\nonumber
&=& \xi_{\rm Ly,T} 
+ \langle\Delta \dot{N}_{\rm Ly} \sigma_{\rm T}\rangle   
+ \langle\Delta \dot{N} \Delta T\rangle + \langle\Delta \dot{N} \sigma_{\rm T}\rangle  \\
&&\hspace{-3mm} 
+ \langle\sigma_{\rm fg} \Delta T \rangle 
+ \langle\sigma_{\rm fg}\sigma_{\rm T}\rangle  
+ \langle\sigma_{\rm flat} \Delta T \rangle 
+ \langle\sigma_{\rm flat}\sigma_{\rm T}\rangle . 
\end{eqnarray}
Here we have assumed prior removal of spectrally smooth foreground from the
redshifted 21cm maps (this removal is expected to be part of the real time
data processing pipeline for an instrument like the LFD).  We have also defined $\Delta T\equiv T-\langle T\rangle$. The fluctuations and noise in the fore-ground should be
un-correlated with the 21cm signal. Similarly, the noise in the 21cm signal
should be uncorrelated with each of the Ly$\alpha$ fluctuations, the noise
in Ly$\alpha$ flux, and the level of foreground. Terms 2-8 in the above
equation therefore average individually to zero over a large
sample. However for a finite sample, the expectation value will have a
distribution with a finite variance about zero. To examine the variance,
consider two variables $x$ and $y$. Their product has a distribution
$p(xy)$ with variance $\sigma_{xy}$. If we sample this distribution $N_{\rm points}$
times, the resulting mean is distributed about zero with a variance
$\langle xy \rangle=\sigma_{xy}/\sqrt{N_{\rm points}}$. Since 21cm surveys are
inherently wide-field, the number of independent terms in the
cross-correlation will be limited by optical surveys. The width of the frequency bin $\Delta\nu_{\rm 21}$ corresponds to the line-of-sight depth of a spherical region of radius $\theta$. At small angles this depth can be smaller than the line-of-sight distance corresponding to a narrow ($100$\AA~) near-IR band. Thus if a map of
Ly$\alpha$ emission has an area $A_{\rm sky}$, then the number of regions
is 
\begin{equation}
N_{\rm points}\sim \frac{A_{\rm sky}}{\pi\theta^2}\left(\frac{\Delta\nu_{\rm Ly}/\nu_{\rm Ly}}{\Delta\nu_{21}/\nu_{21}}\right),
\end{equation}
where $\nu_{21}$ and $\nu_{\rm Ly}$ are the redshifted frequencies of the 21cm and Ly$\alpha$ emission respectively.

Figure~\ref{fig3} shows each of the noise terms in equation~(\ref{CC}) as a function of $\theta$, corresponding to the case of a Ly$\alpha$ survey with an area of $A_{\rm sky}=10$ square degrees flattened at the 1\% level ($f_{\rm flat}=0.01$), with an LFD integration time of 1000 hours. Also shown is the expected cross-correlation function (large dots). In the case shown, the cross correlation would be only marginally detectable. The figure demonstrates that at large angular scales the detection is limited by the flatness of the Ly$\alpha$ field achieved in the experiment. At small angular scales the detection is limited by the error in the brightness temperature of a synthesized 21cm beam. Improved measurements would therefore require flatter fields and larger radio arrays rather than deeper near IR imaging.

The signal-to-noise ratio for detection of the cross-correlation is given by
\begin{equation}
(SN)^2 = \frac{(\xi_{\rm Ly,T})^2}{\Sigma^2}
\end{equation}
where
\begin{eqnarray}
\nonumber 
\Sigma^2 &=& \langle\Delta \dot{N}_{\rm Ly} \sigma_{\rm T}\rangle^2   
+ \langle\Delta \dot{N} \Delta T\rangle^2
+ \langle\Delta \dot{N} \sigma_{\rm T}\rangle^2  \\
&+& \langle\Delta T \sigma_{\rm fg}\rangle^2 
+ \langle\sigma_{\rm flat}\sigma_{\rm T}\rangle^2
+ \langle\sigma_{\rm flat}\sigma_{\rm T}\rangle^2.
\end{eqnarray}
Signal-to-noise ratios as a function of angle are plotted in Figure~\ref{fig4} assuming parameters corresponding to a range of observational facilities. Six cases are shown in each panel, corresponding to Ly$\alpha$ surveys with areas of $A_{\rm sky}=10$ and 100 square degrees performed using a 2m telescope with 1 hour of integration per pointing; combined with low-frequency arrays of collecting area corresponding to 1, 10 and 100 LFDs with an integration time 1000 hours. The latter example of a low-frequency array has around a square
kilometer of collecting area and would represent the realisation of a square kilometer array (see
http://www.skatelescope.org/). The left and right panels correspond to space based (i.e. no atmospheric skyglow, but including zodiacal light), and ground based (i.e. including skyglow) imaging. The 2m space based telescope capable of widefield imaging might represent a telescope like the planned Supernova Acceleration Probe (SNAP\footnote{see http://snap.lbl.gov/}). The value of $f_{\rm flat}$ (ranging between 0 and 0.01) is listed in each case. We have shown examples with values of $f_{\rm flat}$ that are an order of magnitude smaller for ground based examples. The upper row with $f_{\rm flat}=0$ represents an experiment with a perfectly flat near-IR image field.

Figure~\ref{fig4} shows that the cross correlation will be detectable at
angles below a few arc-minutes using wide-field space based imaging
($\sim100$ square degrees) combined with 10 times the LFD, provided that
the Ly$\alpha$ images can be flattened at the $\sim0.1\%$ level. Ground
based studies will be limited by the flatness of the near-IR imaging field,
and would need to reach values of $f_{\rm flat}\sim10^{-4}$ over 100 square
degrees.  At large angles the signal to noise ratio is limited by the value
of $f_{\rm flat}$, and by the area of the survey. However at small angles
the measurement of cross-correlation is limited by the noise in the 21cm
observations, and so the signal-to-noise ratio is proportional to
collecting area of the low-frequency array. The greater sensitivity of an
SKA therefore increases the signal to noise of the detection on scales near
an arc-minute. For example, a signal-to-noise ratio greater than 10 could
be achieved in a 100 square degree survey combined with a ground based
near-IR survey with $f_{\rm flat}=10^{-4}$ or with a space based near-IR
survey with $f_{\rm flat}=0.01$.

\subsection{Signal-to noise ratio in the case of uniform zodiacal light}

The signal-to-noise ratio results presented thus far have assumed that the
sky-glow and zodiacal light leave an imprint on the measured fluctuations
via the instrumental effect of an imperfect flat-field. 
However zodiacal light has very small spatial and temporal fluctuations
(Kashlinsky, Arendt, Mather \& Moseley~2007). In space based observations,
the fluctuations introduced by variable instrumental response to the
zodiacal light could therefore be removed via subtraction of two independent
regions of sky, which we label A and B. Before concluding this paper, we
therefore estimate the signal-to-noise ratio of an analysis conducted this
way.

\begin{figure*}
\includegraphics[width=11.5cm]{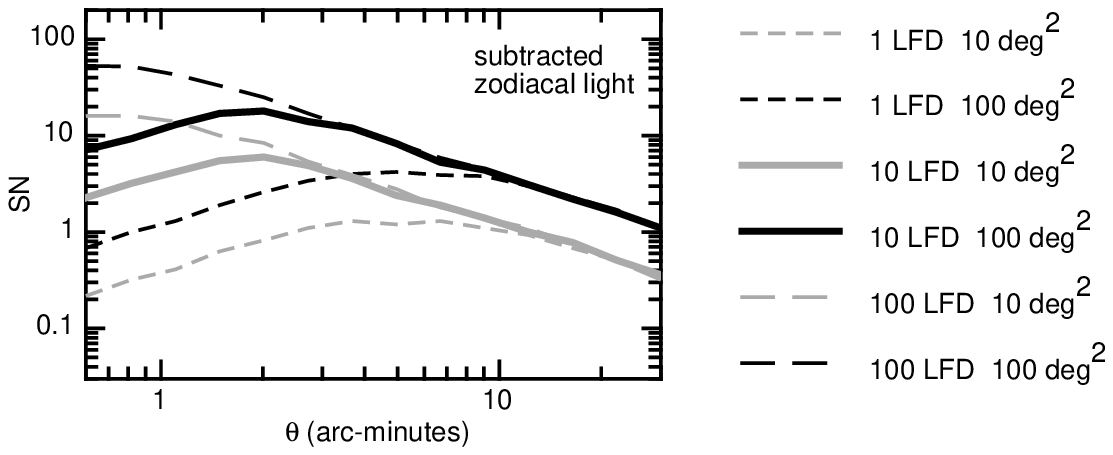} 
\caption{ Signal to noise ratios as a function of angle. Six
cases are shown, corresponding to Ly$\alpha$ surveys with areas of $A=10$
and 100 square degrees using a 2m telescope and a 1 hour integration;
combined with low-frequency arrays of collecting area corresponding to 1, 10
and 100 LFDs with an integration time 1000 hours. The simulation assumes space based observation (i.e. no sky-glow, but including zodiacal light), where the contribution to fluctuations due to imperfections in the flatness of the field are removed via subtraction of the constant fore-ground which is dominated by zodiacal light. }
\label{fig5}
\end{figure*}

As in equation~(\ref{CC}), the observed cross-correlation function
($\xi_{\rm Ly,T}$) is a combination of real fluctuations and noise, hence
we can write the cross-correlation of differences between two regions of
the map, \textit{measured using the same area of detector} as
\begin{eqnarray} 
\label{CC2}
\nonumber
&&\frac{1}{2} \langle \left[ \left(\Delta \dot{N}_{\rm Ly}^{\rm A}+\Delta\dot{N}^{\rm A}+\sigma_{\rm fg}^{\rm A}+\sigma_{\rm flat}\right)\right. \\
\nonumber &&\hspace{20mm} -\left.\left(\Delta \dot{N}_{\rm Ly}^{\rm B}+\Delta\dot{N}^{\rm B}+\sigma_{\rm fg}^{\rm B}+\sigma_{\rm flat}\right)\right] \\
\nonumber &&\hspace{20mm}\times \left[\left(\Delta T^{\rm A} + \sigma_{\rm T}^{\rm A} \right)-\left(\Delta T^{\rm B} + \sigma_{\rm T}^{\rm B} \right)\right] \rangle \\
\nonumber
&=& \xi_{\rm Ly,T} + \langle\Delta \dot{N}_{\rm Ly}^{\rm A} \Delta T^{\rm B}\rangle \\
\nonumber && \hspace {10mm} + 2\left[\langle\Delta \dot{N}_{\rm Ly} \sigma_{\rm T}\rangle   
+ \langle\Delta \dot{N} \Delta T\rangle + \langle\Delta \dot{N} \sigma_{\rm T}\rangle\right.\\
&&\hspace{30mm} \left.+ \langle\sigma_{\rm fg} \Delta T \rangle 
+ \langle\sigma_{\rm fg}\sigma_{\rm T}\rangle\right] 
\end{eqnarray}
In obtaining the second line of equation~(\ref{CC2}), we have noted that if
the zodiacal light is constant then the contribution from terms containing
$\sigma_{\rm flat}$ disappears\footnote{Note that there is still a
contribution to $\sigma_{\rm flat}$ resulting from the extra-galactic
foreground. However this contribution is $\sim100$ times smaller than the
contribution from zodiacal light and we ignore it for this calculation}. We
have also used the fact that the product of un-correlated quantities
between different regions A and B has the same distribution as the product
obtained from the same region (A or B). In this case the signal-to-noise
ratio for detection of the correlation is given by
\begin{equation}
(SN)^2 = \frac{(\xi_{\rm Ly,T})^2}{\Sigma^2}
\end{equation}
where
\begin{eqnarray}
\label{SN2}
\nonumber 
\Sigma^2 &=& \langle\Delta \dot{N}_{\rm Ly}^{\rm A} \Delta_{\rm T}^{\rm B}\rangle^2 + 4\left[\langle\Delta \dot{N}_{\rm Ly} \sigma_{\rm T}\rangle^2 \right.\\  
&&\left.\hspace{10mm}+ \langle\Delta \dot{N} \Delta T\rangle^2
+ \langle\Delta \dot{N} \sigma_{\rm T}\rangle^2 + \langle\Delta T \sigma_{\rm fg}\rangle^2 \right]
\end{eqnarray}

Signal to noise ratios as a function of angle are plotted in
Figure~\ref{fig5} assuming parameters corresponding to a range of
observational facilities. As before six cases are shown, corresponding to
Ly$\alpha$ surveys with areas of $A_{\rm sky}=10$ and 100 square degrees
performed using a 2m telescope with 1 hour integrations; combined with
low-frequency arrays of collecting area corresponding to 1, 10 and 100 LFDs
with an integration time 1000 hours. Only one panel is shown because only
space based observations have been considered, and because the SN
calculated using equation~(\ref{SN2}) is not dependent on the parameter
$f_{\rm flat}$.

Figure~\ref{fig5} shows that at angles below a few arc-minutes, the
cross-correlation will be detectable using space-based wide-field imaging
($\sim10$ square degrees) combined with a low-frequency-array collecting
area of at least 10 times that of the LFD. At large scales the signal to
noise ratio is substantially improved relative to
Figure~\ref{fig4}. Figure~\ref{fig5} shows that space-based near-IR surveys
with areas of 100 square degrees could achieve $SN\sim5$ on angles of
$5-10^\prime$ when combined with the LFD.

In principle, the removal of atmospheric skyglow could also be accomplished through subtraction of independent regions of sky in analogy to equations~(\ref{CC2}-\ref{SN2}).  However since (unlike the zodiacal light) the atmospheric skyglow is variable on short timescales, the removal would need to be averaged over a large number of pointings. We have not attempted to compute the signal-to-noise for a detection in this case.

\section{Discussion}

Standard models for stellar reionization of the IGM predict a clear
anti-correlation between the distribution of bright resolved galaxies and
the 21cm signal (Wyithe \& Loeb~2006). Moreover this anti-correlation would
be easily detectable (Wyithe \& Loeb~2007; Furlanetto \&
Lidz~2007). However reionization is thought to be dominated by low mass
galaxies. In this paper we have demonstrated that the cross-correlation
between fluctuations in the surface brightness of Ly$\alpha$ emission (as a
proxy for star formation rate) and the redshifted 21cm signal, will
directly test the existence of a causal link between the production of
ionizing photons by stars and the reionization of the IGM.

The faint galaxies that make up the unresolved component of high
redshift emission produce most of the Ly$\alpha$ emission (and
corresponding UV radiation). One might therefore suppose that it
should be easier to detect the correlation between the unresolved
component and the 21cm emission.  However (once a redshift is
measured) fluctuations in the resolved galaxy distribution do not
suffer from extragalactic foreground or sky brightness contamination,
while the unresolved emission must be separated from the fluctuating
foreground statistically based on its cross-correlation with
the 21cm signal. The correlation of the 21cm signal with a fluctuating
Ly$\alpha$ surface brightness will therefore be substantially more
difficult to detect than a correlation with resolved galaxies (Wyithe \& Loeb~2007; Furlanetto \& Lidz~2007).

In this paper we have assumed that measurement of the cross-correlation between 21cm
intensity and diffuse Ly$\alpha$ emission would be performed using observations in a narrow
near-IR band. The advantage of a narrow band is that the relative
fluctuations in both Ly$\alpha$ and 21cm emission are larger than they would be if
averaged over a wider line-of-sight interval, corresponding to a broad
band. On the other hand, the signal-to-noise ratio in an individual
pointing is increased if a broad band is used, making detection of the
smaller fluctuations easier. The key ingredient to measuring the
fluctuations due to star formation at high redshift is the ability to
remove fluctuations due to the foreground galaxies. As we have shown, these
fluctuations are comparable in magnitude to the Ly$\alpha$ signal. Existing
measurements of fluctuations in unresolved emission have required
subtraction of an estimated fluctuating foreground component (Kashlinsky et
al.~2005).  Here we suggest removal of the foreground fluctuations
statistically using the fact that these are uncorrelated with the
redshifted 21cm emission. The 21cm emission also has a fluctuating
foreground, and this foreground will be correlated with the foreground in
the Ly$\alpha$ observations. It is proposed as part of upcoming 21cm
experiments, that this redshifted 21cm foreground be removed using the
smoothness of the spectrum of foreground sources (which will be compared
with the rapid frequency fluctuations of the 21cm signal). This 
subtraction method will reveal the narrow-band 21cm
fluctuations, but will not allow detection of broad-band fluctuations. Hence a
narrow-band near-IR observation would be the optimal choice for detecting
the cross-correlation between the Ly$\alpha$ and 21cm signals.

We have analyzed the prospects for detection of the predicted
cross-correlation between Ly$\alpha$ and 21cm emission, and found that
detection will be possible at angular scales smaller than $\sim10^\prime$. At scales of $5-10^\prime$ the measurement could be performed
using near-IR imaging from space, combined with a low frequency array
having a collecting area equal to that of the LFD. At smaller angular
scales the SN can be significantly increased, but will require a collecting
area for redshifted 21cm observations of at least 10 times the
LFD. Observations from the ground will be limited by the difficulties of
subtracting a sufficiently flat sky, which would be dominated by
atmospheric sky-glow.  When 21cm observations are combined with space-based
near-IR imaging, the detection on scales smaller than a few arc-minutes
will be limited by the sensitivity to the 21cm signal. This will be true
even when the experiment is performed with a small aperture optical
telescope over a moderate field of view ($\sim10$ square degrees). 

A futuristic wide-field space-based survey telescope combined with a
Square-Kilometer-Array would detect the cross-correlation at very high
signal-to-noise ratios over a range of angular scales below a few
arc-minutes.  The space-based near-IR survey need not have a very highly sampled
point-spread function beyond that necessary for the subtraction of resolved
galaxies. A space based survey telescope like the proposed Supernova
Acceleration Probe (SNAP) would therefore provide the ideal facility with
which to explore the connection between star formation and the reionization
of the universe. To perform an experiment of the sort proposed in this paper, the survey telescope would need to carry an appropriate narrow-band filter. To study star formation at $z\sim7$, corresponding to the examples presented in this paper the filter should be centered at a wavelength of $\sim 9700\AA$ with a width of $\sim100\AA$.

{\bf Acknowledgments} The research was supported by the Australian Research
Council (JSBW \& BPS) and Harvard University grants (AL). JSBW acknowledges the hospitality of the Institute of Astronomy at Cambridge University where this work was completed.

\newcommand{\noopsort}[1]{}

\label{lastpage}
\end{document}